\pdfoutput=1
\newcommand*{\ATLASLATEXPATH}{}
\documentclass[USenglish,block=none,cernpreprint,texlive=2016]{\ATLASLATEXPATH atlasdoc}
 
\usepackage{\ATLASLATEXPATH atlaspackage}
\usepackage{\ATLASLATEXPATH atlasbiblatex}
 
\usepackage{\ATLASLATEXPATH atlasphysics}
 
\graphicspath{{logos/}{figures/}}
 
\usepackage{ANA-STDM-2017-06-PAPER-defs}
 
\usepackage{placeins}
\usepackage{multirow}
\usepackage{pdflscape}
\usepackage{float}
\usepackage{graphicx} 
\usepackage{booktabs} 
 
\AtlasTitle{Observation of electroweak production of a same-sign $W$ boson pair in association with two jets in $pp$ collisions at $\sqrt{s}=13$~\TeV\ with the ATLAS detector}
 
\author{The ATLAS Collaboration}
 
\AtlasRefCode{STDM-2017-06}
\AtlasNote{STDM-2017-06}
 
\PreprintIdNumber{CERN-EP-2019-008}
 
\AtlasJournalRef{Phys.\ Rev.\ Lett.\ 123 (2019) 161801}
\AtlasDOI{10.1103/PhysRevLett.123.161801}
 
\AtlasAbstract{
This Letter presents the observation and measurement of electroweak production of a same-sign $W$ boson pair in association with two jets using 36.1~fb$^{-1}$ of proton--proton collision data recorded at a center-of-mass energy of $\sqrt{s}=13$~\TeV\ by the ATLAS detector at the Large Hadron Collider. The analysis is performed in the detector fiducial phase-space region, defined by the presence of two same-sign leptons, electron or muon, and at least two jets with a large invariant mass and rapidity difference. A total of 122 candidate events are observed for a background expectation of $69 \pm 7$ events, corresponding to an observed signal significance of 6.5 standard deviations. The measured fiducial signal cross section is $\sigma^{\textrm {fid.}}=2.89^{+0.51}_{-0.48}~\textrm{(stat.)}\ ^{+0.29}_{-0.28}~\textrm{(syst.)}\textrm{ fb}$.
}
 
\hypersetup{pdftitle={ATLAS draft},pdfauthor={The ATLAS Collaboration}}
 
\begin{document}
 
\maketitle
 
The scattering of two massive vector bosons (VBS), $VV \rightarrow VV$ with $V=W$ or $Z$, is an important process for studying the mechanism of electroweak symmetry breaking~\cite{Chanowitz:1988ft, Bagger:1993zf, Bagger:1995mk}. VBS processes involve quartic gauge-boson self interactions, and the $s$- and $t$-channel exchanges of a gauge or Higgs boson.  The Higgs boson regularizes the VBS amplitude by canceling out the divergencies arising from longitudinally polarized vector bosons at high energy~\cite{Lee:1977yc, Lee:1977eg}. These cancellations depend on the gauge structure of the theory and are exact in the Standard Model (SM)~\cite{tHooft:1972tcz,Veltman:1976rt}. The present measurement of $W$ boson scattering thus serves as a fundamental probe of the SM electroweak theory.
 
\begin{figure}[htb]
\centering
\subfloat[]{
\includegraphics[width=.245\textwidth]{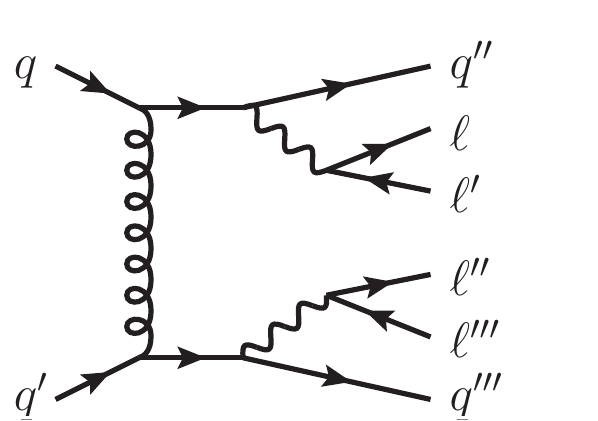}
\label{fig:feyn:qcd}
}
\subfloat[]{
\includegraphics[width=.245\textwidth]{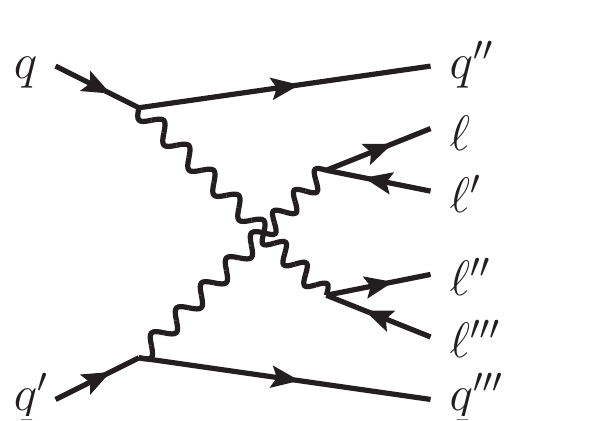}
\label{fig:feyn:qgc}
}
\caption{
Representative diagrams for $VVjj$ production where two electroweak gauge bosons are radiated off quarks. (a) Strong production diagrams involve both electroweak and strong interactions without gauge boson self interactions. (b) In VBS diagrams, the gauge bosons can interact via gauge-boson self interactions, e.g.~a quartic $W$ boson vertex.
}
\end{figure}

At the LHC, the VBS final state of two gauge bosons and two jets ($VVjj$) can be produced via two classes of mechanisms.  The first class, referred to as strong production, involves both strong and electroweak interactions at Born level, and features diagrams where the incoming partons exchange color, as illustrated in Figure~\ref{fig:feyn:qcd}. The second class, referred to as electroweak production, involves only weak interactions at Born level~\cite{Biedermann:2017bss} and includes VBS diagrams. Figure~\ref{fig:feyn:qgc} shows a typical VBS diagram where the gauge bosons are radiated off the incoming quarks and then scatter via the quartic self interaction vertex. In VBS processes, the incoming partons do not exchange color and typically produce the two jets with a large invariant mass and with large rapidity difference~\cite{Ballestrero:2018anz}.
 
The $\sswwjj$ final state has the largest ratio of electroweak to strong production cross sections compared to other VBS diboson processes~\cite{Bagger:1995mk}; this is because at leading-order (LO) accuracy in perturbative quantum chromodynamics (QCD) quark-gluon and gluon-gluon initiated diagrams are absent and contributions from quark and (anti-)quark annihilation diagrams are suppressed. This ratio is of order five in the fiducial phase-space region of this analysis. The $s$-channel VBS diagrams with trilinear self interactions are absent in this final state. In addition, electroweak diagrams not involving self interactions are suppressed~\cite{Szleper:2014xxa}, thus enhancing sensitivity of this final state to gauge-boson self couplings. Previously, an observation of $\sswwjj$ electroweak production was reported by the CMS Collaboration~\cite{Sirunyan:2017ret} and evidence was reported by the ATLAS Collaboration using a smaller dataset~\cite{STDM-2013-06, STDM-2014-05}.
 
This Letter presents the observation and measurement of the electroweak production of \sswwjj\ events in which both $W$ bosons decay into an electron or muon and a neutrino. This study uses {\lumi} of proton--proton ($pp$) collision data collected by the ATLAS detector at $\sqrt{s}=13$~\TeV. The ATLAS detector~\cite{PERF-2007-01} is a multipurpose particle detector with a forward--backward symmetric cylindrical geometry and almost $4\pi$ coverage in solid angle.\footnote{ATLAS uses a right-handed coordinate system with its origin at the nominal interaction point (IP) in the center of the detector and the $z$-axis along the beam direction. The $x$-axis points from the IP to the center of the LHC ring, and the $y$-axis points upward. Cylindrical coordinates $(r,\phi)$ are used in the $(x, y)$ plane, $\phi$ being the azimuthal angle around the beam direction. The rapidity is defined as $y=\frac{1}{2}\ln \left( \frac{E+p_z}{E-p_z} \right)$ where $E$ is the energy of the particle and $p_z$ is the projection of the momentum along the $z$-axis.  The pseudorapidity is defined in terms of the polar angle $\theta$ as $\eta=-\ln\tan(\theta/2)$. The distance $\Delta R$ is defined as $\Delta R=\sqrt{(\Delta \eta)^2+(\Delta \phi)^2}$.} The inner tracking detector (ID) covers $|\eta| < 2.5$ in pseudorapidity and consists of a silicon  pixel detector, a silicon microstrip detector, and a transition radiation tracker. The ID is surrounded by a superconducting solenoid magnet and an almost hermetic calorimeter system, which provides three-dimensional reconstruction of particle showers up to $|\eta| = 4.9$. The muon spectrometer (MS) has three air-core toroidal magnets: a barrel toroid and two endcap toroids. Three layers of precision tracking stations with drift tubes and cathode strip chambers allow precise muon momentum measurement up to $|\eta| = 2.7$. Resistive-plate and thin-gap chambers provide muon triggering capability up to $|\eta| = 2.4$.
 
Monte Carlo (MC) simulation is used to generate signal and background predictions. The generated events are processed through a detector simulation~\cite{SOFT-2010-01} based on {\textsc {Geant4}}~\cite{Agostinelli:2002hh} and are reconstructed using the same algorithms as used for data. The simulation includes the effect of multiple $pp$ interactions per bunch crossing  as detailed in Ref.~\cite{Aaboud:2016lpj}.
 
Processes producing four leptons were simulated using the event generator {\sherpa}~v2.2.2~\cite{Gleisberg:2008ta, Gleisberg:2008fv, Schumann:2007mg, Hoeche:2009rj} with the NNPDF3.0NNLO \cite{Ball:2014uwa} set of parton distribution functions (PDF). Electroweak diboson production in association with two jets was simulated with diagrams including exactly six orders of the EW coupling~\cite{ATL-PHYS-PUB-2019-004}. The simulation of the strong production processes includes diagrams with exactly four electroweak (EW) vertices~\cite{ATL-PHYS-PUB-2017-005}. The simulation of both the \sswwjj\ electroweak signal and \sswwjj\ strong background processes includes up to one additional parton at LO in QCD. For $\ell\ell\ell\nu$ and 4$\ell$ final states, matrix elements include up to one parton at next-to-leading-order (NLO) and three partons at LO in QCD. All multileg processes were simulated combining the various final-state topologies with the MEPS merging algorithm~\cite{Hoeche:2012yf, Hoeche:2009rj} and matched to the {\sherpa} internal parton shower, hadronization and underlying-event modeling~\cite{Schumann:2007mg, Schonherr:2008av}. The renormalization and factorization scales were set to the invariant mass of the four-lepton system at the matrix-element level. A signal sample for alternative theoretical predictions was simulated with \POWHEGBOX~\cite{Jager:2011ms, Nason:2004rx, Frixione:2007vw, Alioli:2010xd} and {\pythia.230}~\cite{Sjostrand:2014zea} at NLO QCD accuracy and employing the VBS approximation~\cite{Jager:2009xx}, the NNPDF3.0NLO PDF set, and renormalization and factorization scales set to the $W$ boson mass. The $V\gamma$ processes were simulated with exactly three EW vertices using {\sherpa}~v2.1.1 at NLO (up to one parton) or LO accuracy (up to three partons) using the CT10NLO PDF set~\cite{Lai:2010vv}. Electroweak $V\gamma jj$ processes were simulated using {\sherpa}~v2.2.4 with the NNPDF3.0NNLO PDF set, with exactly five orders of the EW coupling, and at LO QCD accuracy. The triboson production processes were simulated with {\sherpa}~v2.1.1  and the CT10NLO PDF set. {\textsc {Madgraph5\_aMC@NLO}}~\cite{Alwall:2014hca} with the NNPDF3.0NLO PDF set and {\pythia.210} were used to simulate $t\bar{t}V$ processes at NLO QCD accuracy.
 
The $\sswwjj$ electroweak production cross section is measured in a fiducial region defined at particle level in MC simulation by requiring exactly two same-sign leptons with a transverse momentum, $\pt$, greater than 27~\GeV\ and $|\eta|<2.5$; leptons are defined as electrons and muons produced in $W$ boson decays and that do not originate from $\tau$ decays. The lepton four-momentum includes the four-momenta of photons inside a cone of size $\Delta R = 0.1$ around the lepton. The two leptons must be separated by a distance of $\Delta R_{\ell \ell}>0.3$ and must have an invariant mass, $\mll$, greater than 20~\GeV. The magnitude of the vector sum of the transverse momenta of the two final-state neutrinos with highest $\pt$ must be greater than 30~\GeV. Jets are reconstructed using the anti-$k_t$ algorithm~\cite{Cacciari:2008gp} with radius parameter $R = 0.4$ and using all final-state particles, except for neutrinos and charged leptons from $W$ boson decays. Jets are required to have $\pt > 35$~\GeV\ and $|\eta|<4.5$. Events with a charged lepton that is within a cone of radius $\Delta R_{\ell j} = 0.3$ around a jet are vetoed. The fiducial region requires at least two jets, including one with $\pt>65$~\GeV\ and another with $\pt>35$~\GeV. The two highest-$\pt$ jets must have an invariant mass $\mjj > 500$~\GeV\ and a rapidity difference $|\dyjj|>2$.
 
The fiducial cross section predicted by {\sherpa} for $\sswwjj$ electroweak production is $\shpxsec$. The uncertainty includes independent variations of the renormalization and factorization scales by factors of 0.5 and 2 with the constraint $0.5 \leq \mu_{\mathrm{F}}/\mu_{\mathrm{R}} \leq 2$ which contribute $^{+14\%}_{-11\%}$; it also includes uncertainties from the NNPDF3.0NNLO ensemble, as well as differences between the CT14~\cite{Dulat:2015mca} and MMHT2014~\cite{Harland-Lang:2014zoa} PDF sets ($^{+2.5\%}_{-1.5\%}$)~\cite{Butterworth:2015oua}. Uncertainties in the parton shower, hadronization, and underlying-event modeling are evaluated by varying the MEPS matching and resummation scales and amount to $^{+8\%}_{-1\%}$. {\powheg}+{\pythia} predicts a signal fiducial cross section of $\pbpxsec$, with the uncertainties derived using the same procedures as for the {\sherpa} prediction, except for the uncertainty in the parton shower modeling, which is estimated as the difference relative to {\powheg}+\textsc{Herwig7}~\cite{Bahr:2008pv,Bellm:2015jjp}. The {\sherpa} electroweak samples suffer from a non-optimal setting of the color flow, which leads to an excess of central jet emissions from the parton shower. Since up to one additional parton is included in the matrix element of the $\sswwjj$ electroweak sample, the effect on its differential distributions is reduced but accompanied by a significant suppression of the predicted cross section~\cite{ATL-PHYS-PUB-2019-004}.
 
Events are required to contain at least one reconstructed proton interaction vertex. The vertex with the highest $\pt^2$ sum of associated ID tracks is selected as the primary vertex. Electrons are reconstructed from energy clusters in the electromagnetic calorimeter that are matched to tracks reconstructed in the ID with the requirement of a hit in the innermost pixel layer~\cite{PERF-2017-01}. Muon are reconstructed by combining ID and MS information \cite{PERF-2015-10}. Electron and muon candidates must satisfy loose identification criteria~\cite{PERF-2017-01,PERF-2015-10}, have $\pt > 6~\GeV$, and $|\eta| < 2.47$ and $|\eta| < 2.7$, respectively. The ID tracks associated to electron (muon) candidates are matched  to the primary vertex by requiring their transverse impact parameter significance to satisfy $|d_{0}/\sigma_{d_0}|<5~(10)$; the longitudinal impact parameter multiplied by the sine of the polar angle of the lepton candidates must satisfy $|z_{0}  \sin\theta|<0.5$~mm. Electrons and muons passing these selections are further referred to as baseline leptons.
 
Jets are reconstructed from calorimeter energy clusters~\cite{PERF-2014-07,PERF-2016-04} using the anti-$k_t$ algorithm with radius parameter $R = 0.4$. Jets are required to have $\pt>30$~\GeV\ in the forward region ($2.4<|\eta|<4.5$) and $\pt>25$~\GeV\ in the central region ($|\eta|<2.4$). Central jets with $\pt<60$~\GeV\ must be matched to the primary vertex~\cite{PERF-2014-03}. Jets containing $b$-hadrons ($b$-jets) are identified in the range of $|\eta|<2.5$  with an efficiency of 85\% using techniques described in Ref.~\cite{Aaboud:2018xwy}. Selected electron, muon and jet candidates are required to be non-overlapping using the procedures described in Ref.~\cite{Aaboud:2016lpj}. The missing transverse momentum, \met, is computed using selected electrons, muons and jets, and the track-based soft term defined in Ref.~\cite{Aaboud:2018tkc}.
 
Events are selected online by single-electron or single-muon triggers~\cite{Aaboud:2016leb}. Candidate events are selected by requiring exactly two same-sign baseline leptons, electron or muon, with ${\mll>20}$~\GeV\ and by requiring ${\met>30}$~\GeV. They are required to contain at least two jets, including one with $\pt>65$~\GeV\ and another with $\pt>35$~\GeV. Events with at least one identified $b$-jet are rejected in order to reduce background contributions from top-quark pair production ($t\bar{t}$). The two highest-$\pt$ jets are required to have ${\mjj>200}$~\GeV\ and ${|\dyjj|>2}$. These jet selection criteria were optimized to separate the $\sswwjj$ electroweak process from the strong production and other background processes.
 
After these selections, the dominant source of background events is due to leptons originating from decays of heavy-flavor hadrons and jets misidentified as electrons, collectively referred to as non-prompt leptons. Additional selection criteria are applied to reduce their contributions. Signal electrons are required to satisfy tight identification criteria~\cite{PERF-2017-01}, to have ${\pt > 27}$~\GeV, and to be outside the calorimeter transition region (${1.37 < |\eta| < 1.52}$). Signal muons are required to satisfy medium identification criteria~\cite{PERF-2015-10}, and to have ${\pt>27}$~\GeV and $|d_{0}/\sigma_{d_0}|<3$. Signal electrons and muons are further required to be isolated from nearby particles, with isolation criteria defined using calorimeter clusters and ID tracks. These isolation criteria are optimized to have an efficiency of at least 90\% for ${\pT>25}$~\GeV\ and at least 99\% for ${\pT>60}$~\GeV~\cite{PERF-2015-10,PERF-2017-01}. For dielectron events, the electron pseudorapidity is restricted to ${|\eta|<1.37}$ and events with {\zpeak} are discarded. These criteria reduce the background from electron charge misreconstruction described later. Candidate events with exactly two signal leptons are said to pass the full event selection.
 
The contributions from the $\wzjj$, $\vgjj$, $\zzjj$, $\ttv$ and triboson production are estimated using simulation. The predicted event yields of the $\wzjj$ and $\vgjj$ processes are normalized to data in dedicated control regions. The normalization of the $\wzjj$ background is determined using events with exactly three baseline leptons, two of which are required to pass the signal lepton selection, and that satisfy the dijet and $\met$ selection criteria. Events from $\vgjj$ production enter the signal region when a photon is misidentified as an electron. The modeling of this misidentification process in simulation is corrected using $Z \rightarrow \mu^+ \mu^- \gamma$ events where a photon is emitted by a muon and then misidentified as an electron. These events are selected by requiring exactly two opposite-sign signal muons, one signal electron, $\met<30$~\GeV\ and a trilepton invariant mass satisfying $75~\GeV < m_{\mu\mu e} < 100~$\GeV. A normalization factor of $1.8$ is derived from this control region and used to correct the simulated $\vgjj$ events. To account for the differences between the $Z\gamma$ and $\wgjj$ processes, the full effect of this correction factor is assigned as a systematic uncertainty, corresponding to $44\%$ of the estimated $\vgjj$ yield. The relative contributions from electroweak and strong production of $\wzjj$ and $\vgjj$ processes are estimated from simulation since this analysis is not sensitive to their different admixtures. Theoretical uncertainties in the predictions of the $\zzjj$, $V\gamma$, triboson, and $\ttv$ backgrounds vary from $20\%$ to $30\%$~\cite{ATL-PHYS-PUB-2017-005,ATL-PHYS-PUB-2016-002,deFlorian:2016spz}.
 
Background contributions with non-prompt leptons are estimated by weighting data events from dedicated control regions by scale factors. These scale factors are measured in dijet events containing exactly one lepton that is $\pt$-balanced by a $b$-jet. The $b$-jet requirement enhances non-prompt lepton contributions and suppresses contributions from $W/Z$ bosons, which are subtracted from data using simulation. The scale factor is defined as the ratio of the number of signal leptons to the number of leptons passing a dedicated background selection. The background leptons are required to pass the baseline lepton selection and fail the signal lepton selection, where background electrons are in addition required to satisfy medium identification criteria~\cite{PERF-2017-01}. Moreover, the background electron (muon) $\pt$ is required to be greater than $20~(15)~\GeV$. Separate scale factors are computed for muons and for central and forward electrons. In order to reduce the dependence on the underlying $\pt$ spectrum of $b$-jets that produce non-prompt leptons, the scale factors are measured as a function of the scalar sum of the background lepton $\pt$ and the additional activity around the lepton. This activity, $\pt^{\textrm{iso30}}$, is quantified by the sum of the $\pt$ of ID tracks that are within a cone of size $\Delta R=0.3$ around the lepton and originate from the primary vertex.
 
Data events, that are weighted by the scale factors, are taken from control regions defined using the full event selection criteria except that one lepton is required to pass the background lepton selection and its $\pt$ is replaced with $\pt+ \pt^{\textrm{iso30}}$, with this sum required to be greater than $27$~\GeV. A statistically independent control region is defined for each bin of the $\mjj$ distribution. The uncertainty of the estimated non-prompt background yields is approximately 50\% in $\mum$ final states and varies between 40\% and 90\% for $\elel$ and $\elmu$ final states. It includes the systematic uncertainty of the scale factors and the statistical uncertainty of the control regions. The former uncertainty is derived from variations in the composition of the dijet control regions where these factors are measured, obtained by varying the selection criteria. The entire method is validated in regions enriched with non-prompt leptons from $t\bar{t}$ ($W+$jet) events selected by requiring exactly two same-sign leptons and exactly one (zero) $b$-jet among a total of at least (less than) two jets. In these regions, the number of observed data events and the number of predicted background events agree within their uncertainties.
 
Opposite-sign lepton pairs pass the full event selection when an electron undergoes an interaction with the detector material resulting in incorrect charge reconstruction. The probability of this charge misreconstruction, $\epsilon_{\textrm{misrec.}}$, is measured in $Z \rightarrow e^+ e^-$ events~\cite{PERF-2017-01} and it increases from about 0.1\% in the central region to a few percent for $|\eta|>2$. The background contributions from electron charge misreconstruction are estimated from data using opposite-sign lepton pairs that satisfy the full event selection criteria, except for the same-sign requirement; these events are weighted by  $\epsilon_{\textrm{misrec.}}$ and the electron energy loss due to the material interaction is corrected with $\eta$-dependent factors derived from simulation~\cite{PERF-2017-01}. The overall method is validated by comparing the number of observed same-sign electron pairs having {\zpeak} with the predicted background yield, with the two numbers agreeing within the systematic uncertainty of $15\%$. This uncertainty is dominated by the statistical uncertainty in the measurement of $\epsilon_{\textrm{misrec.}}$, which is less than $10\%$ for $|\eta|>2$ and up to $20\%$ in the central region. The charge misreconstruction of muons is found to be negligible.
 
The detector systematic uncertainties arising from the mismodeling of the reconstructed objects are estimated primarily from data and their impact on the analysis is assessed using simulated events. The dominant source is the uncertainty of the jet energy scale, which amounts to $2\%$ for the signal and  $10\%$ for the $\wzjj$ background. The uncertainty in the measurement of the integrated luminosity is $2.1\%$~\cite{ATLAS-CONF-2019-021}.
 
The theory modeling uncertainties of the $\mjj$ distributions predicted by {\sherpa} for $\sswwjj$ and $\wzjj$ processes are evaluated using the procedures described above. They account for uncertainties in the total cross section, the acceptance of the fiducial selection, the modeling of the event selection efficiency and the shape of the $\mjj$ distribution. Only the latter two affect the measured fiducial cross section of the $\sswwjj$ signal, since absolute normalization uncertainties cancel in this measurement. The uncertainty in the modeling of the event selection efficiency also accounts for extrapolations from the fiducial phase space to the detector level, in particular for the $\eta$ acceptance in dielectron events. Effects of the NLO electroweak corrections~\cite{Biedermann:2016yds} and of the interference between electroweak and strong $\sswwjj$ production~\cite{Ballestrero:2018anz} are assigned as an uncertainty in the $\mjj$ shape of the $\sswwjj$ signal, amounting to $6\%$ and $4\%$, respectively. This approach is used because  no event generator implemented the complete NLO calculation  until recently~\cite{Chiesa:2019ulk} and because the interference contribution is defined only at the leading order~\cite{Biedermann:2017bss}. The overlap of the photon radiation in the {\sherpa} parton shower model with the NLO EW corrections is found to be negligible.
 
Signal events are categorized by their lepton flavor and charge into six mutually exclusive channels: \elel, \elmu, and \mum, in order to exploit their different signal and background compositions. The signal region is defined as ${\mjj>500~\GeV}$ and further split into four $\mjj$ bins, optimized to increase the expected signal sensitivity. Events with ${200~\GeV < \mjj < 500~\GeV}$ serve as additional control regions, dominated by contributions from non-prompt leptons and $\wzjj$ backgrounds. The resulting 30 bins of the $\mjj$ distributions in the signal and control regions are combined in a profile likelihood fit~\cite{Cranmer:2012sba} to extract the fiducial cross section.
 
The signal strength, a free parameter in the fit, multiplies the expected fiducial $\sswwjj$ electroweak production cross section used to produce the signal template. The signal template of reconstructed $\sswwjj$ electroweak events also includes candidate events with electrons and muons produced in $W$ decays into $\tau$ lepton. Since the fiducial cross section prediction does not include such events, their fractional contribution predicted by the simulation is removed from the fiducial cross section measurement. Systematic uncertainties are included in the fit as nuisance parameters constrained by Gaussian functions. The $\wzjj$ control region is also included in the fit as a single bin and the normalization of the $\wzjj$ background is included as a free parameter. The analysis choices maximize the expected significance for the {\sswwjj} electroweak signal predicted by {\sherpa} at 4.4$\sigma$. A significance of 6.5$\sigma$ is expected by the alternative signal sample simulated with \POWHEGBOX.
 
\begin{figure}[htb]
\begin{center}
\includegraphics[width=0.49\textwidth]{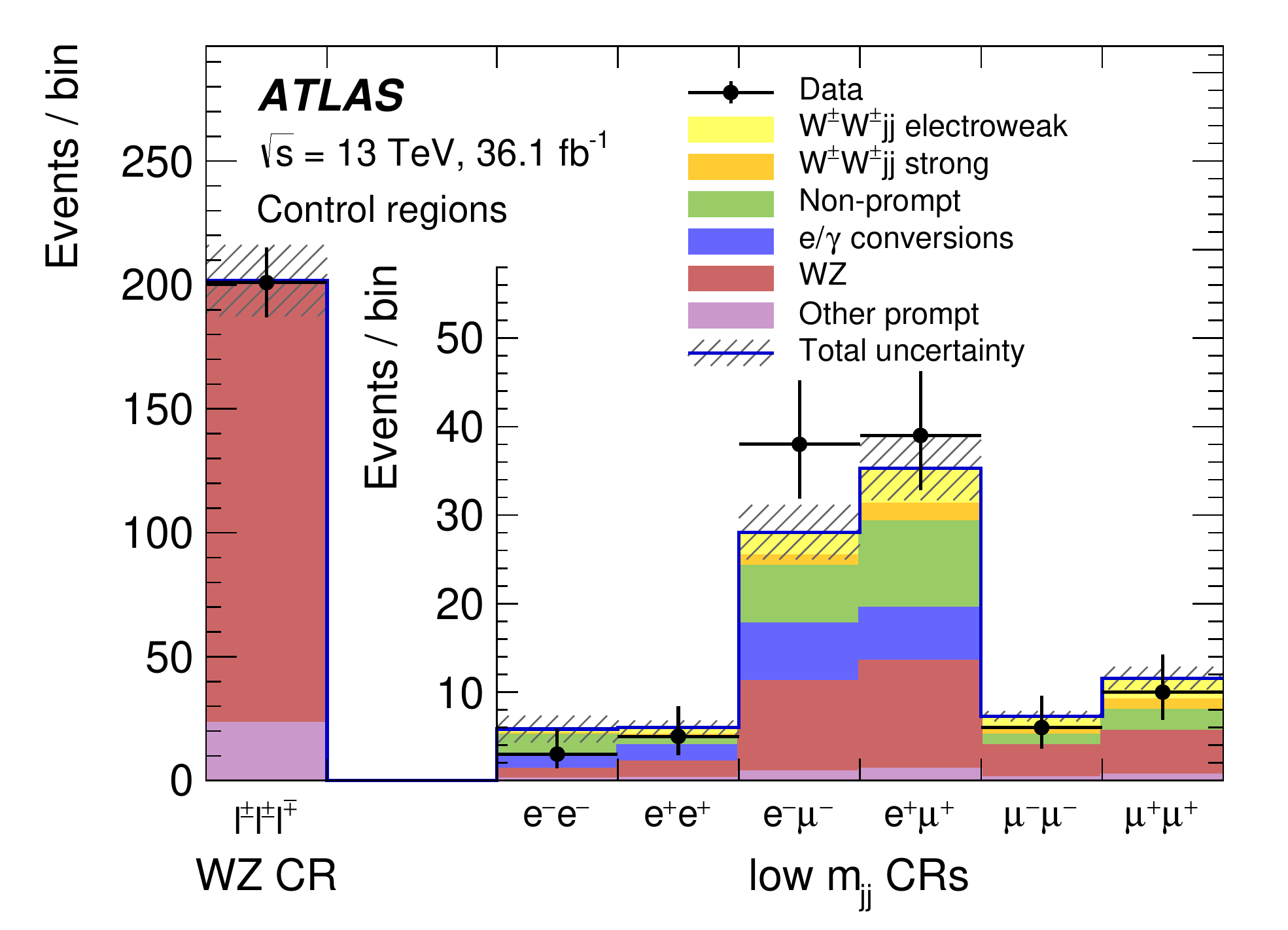}
\includegraphics[width=0.49\textwidth]{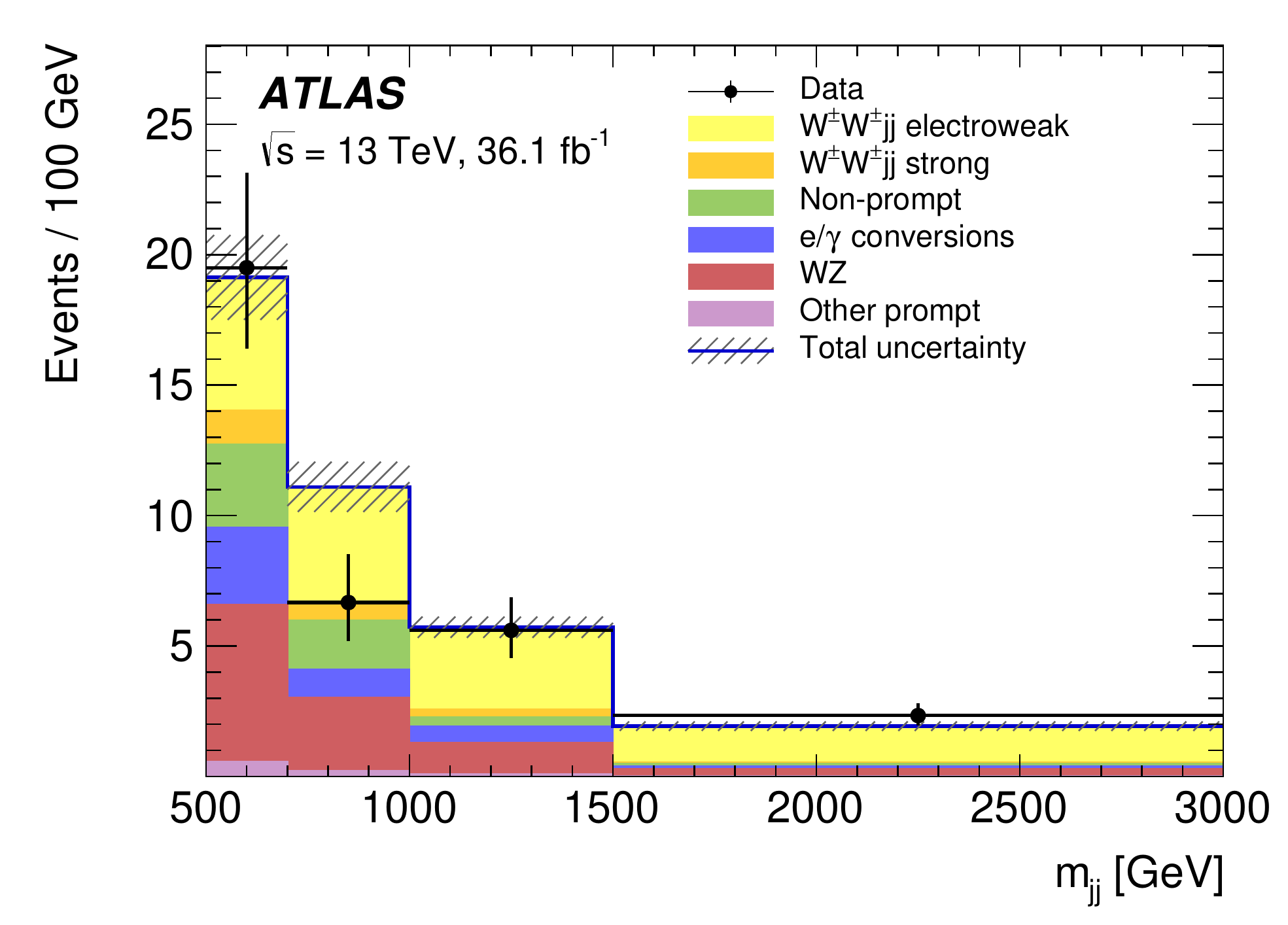}
\caption{Event yields for data, signal and background in the $WZ$ and $200<\mjj<500$~\GeV\ control regions (left) and the $\mjj$ distribution for events meeting all selection criteria for the signal region (right). Signal and background distributions are shown as predicted after the fit. The hatched band represents the statistical and systematic uncertainties of the background predictions added in quadrature. The backgrounds from  $V\gamma$ production and electron charge misreconstruction are combined into the $e/\gamma$ conversions category. The other prompt category combines $ZZ$, $VVV$ and $t\bar{t}V$ background contributions. The last bin on the right figure includes the overflow.}
\label{fig:mjj}
\end{center}
\end{figure}
 
\begin{figure}[htb]
\begin{center}
\includegraphics[width=.49\textwidth]{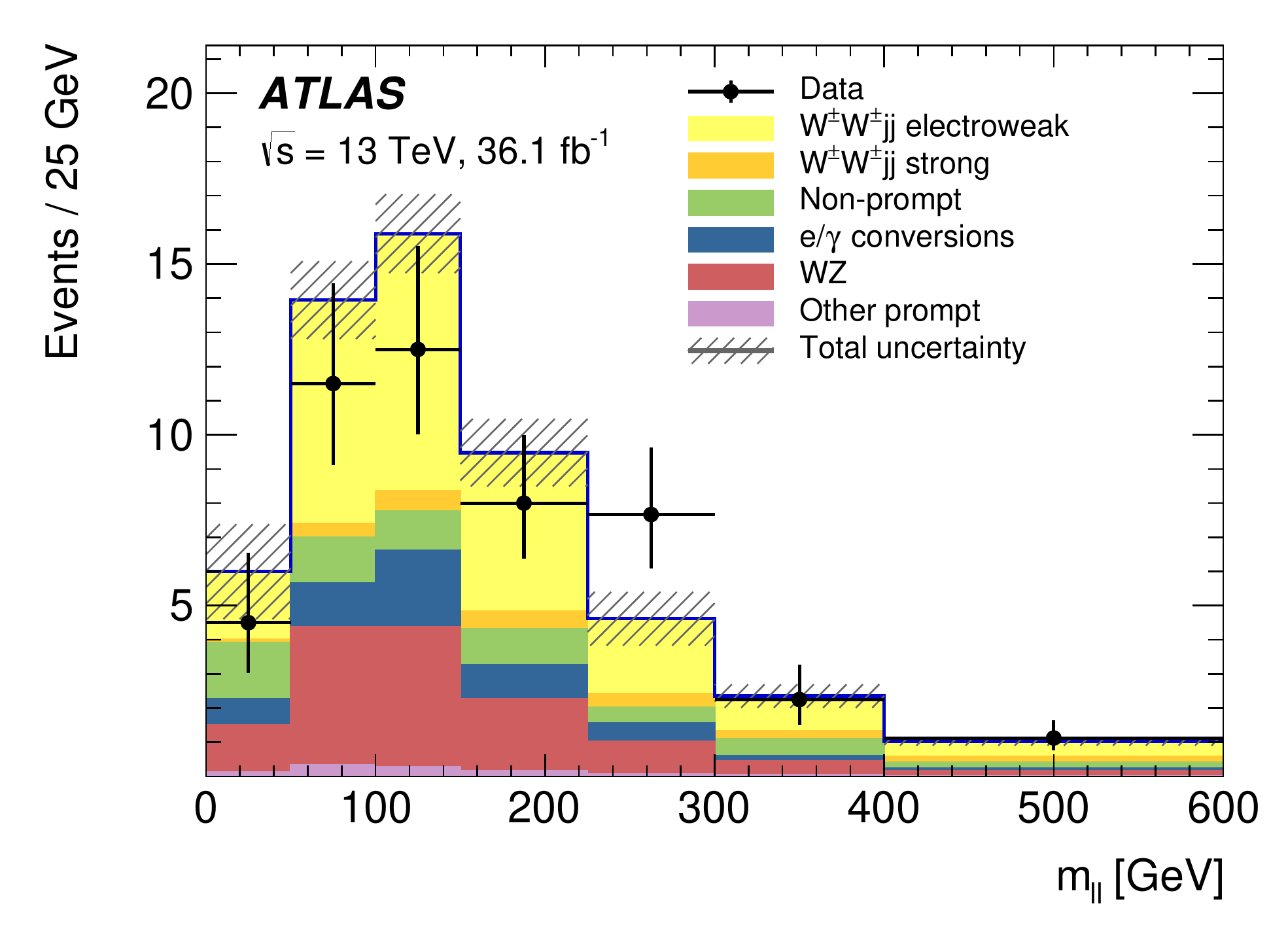}
\caption{
The $\mll$ distribution for events meeting all selection criteria for the signal region is shown as predicted after the fit. The hatched band represents the statistical and systematic uncertainties of the background prediction added in quadrature. The fitted signal strength and nuisance parameters have been propagated, with the exception of the uncertainties due to the interference and electroweak corrections for which a flat uncertainty is assigned. The backgrounds from  $V\gamma$ production and electron charge misreconstruction are combined in the $e/\gamma$ conversions category. The other prompt category combines $ZZ$, $VVV$ and $t\bar{t}V$ background contributions. The last bin of the distribution includes the overflow.}
\label{fig:mll}
\end{center}
\end{figure}

Table~\ref{tab:yields_obs} compares the numbers of data events in the signal region with the background and signal event yields after the fit; the signal region contains 122 data events, compared with a best-fit yield of $69 \pm 7$ background events. By fitting the data and background events in the signal and control regions, the background-only hypothesis is rejected with a significance of 6.5$\sigma$. Figure~\ref{fig:mjj} shows the control region events separated into categories and the $\mjj$ distribution in the signal region after the fit. All nuisance parameters remain within their one standard deviation uncertainty after the fit. The normalization of the $\wzjj$ background is scaled by a factor of \muWZ, constrained mainly by the observed number of data events in the $\wzjj$ control region. Figure~\ref{fig:mll} shows the $\mll$ distribution in the signal region after the fit.
 
\begin{table}[htb]
\caption{Summary of the data event yields, and the signal and background event yields in the signal region as obtained after the fit. The numbers are shown for the six individual channels and for all channels combined. The backgrounds from $V\gamma$ production and electron charge misreconstruction are combined in the $e/\gamma$ conversions category. The other prompt category combines $ZZ$, $VVV$ and $t\bar{t}V$ background contributions.}
\resizebox{\linewidth}{!}{
\resizebox{\linewidth}{!}{
\begin{tabular}{lS@{$\,\pm\,$}SS@{$\,\pm\,$}SS@{$\,\pm\,$}SS@{$\,\pm\,$}SS@{$\,\pm\,$}SS@{$\,\pm\,$}SS@{$\,\pm\,$}S}
\toprule
& \multicolumn{2}{c}{\epep}& \multicolumn{2}{c}{\emem}& \multicolumn{2}{c}{\epmp}& \multicolumn{2}{c}{\emmm}& \multicolumn{2}{c}{\mpmp}& \multicolumn{2}{c}{\mmmm}& \multicolumn{2}{c}{Combined}\\
\midrule
$WZ$                  & 1.48& 0.32   & 1.09& 0.27   & 11.6& 1.9   & 7.9& 1.4   & 5.0& 0.7   & 3.4& 0.6   & 30& 4 \\
Non-prompt            & 2.2& 1.1   & 1.2& 0.6   & 5.9& 2.5   & 4.7& 1.6   & 0.56& 0.05   & 0.68& 0.13   & 15& 5 \\
$e/\gamma$ conversions           & 1.6& 0.4   & 1.6& 0.4   & 6.3& 1.6   & 4.3& 1.1   & \multicolumn{2}{c}{---} & \multicolumn{2}{c}{---}   & 13.9& 2.9 \\
Other prompt          & 0.16& 0.04   & 0.14& 0.04   & 0.90& 0.20   & 0.63& 0.14   & 0.39& 0.09   & 0.22& 0.05   & 2.4& 0.5 \\
\sswwqcd              & 0.35& 0.13   & 0.15& 0.05   & 2.9& 1.0   & 1.2& 0.4   & 1.8& 0.6   & 0.76& 0.25   & 7.2& 2.3 \\
\midrule
Expected background   & 5.8& 1.4   & 4.1& 1.1   & 28& 4   & 18.8& 2.6   & 7.7& 0.9   & 5.1& 0.6   & 69& 7 \\
\midrule
\sswwew            & 5.6& 1.0   & 2.2& 0.4   & 24& 5   & 9.4& 1.8   & 13.4& 2.5   & 5.1& 1.0   & 60& 11 \\
\midrule
Data& \multicolumn{2}{c}{10}& \multicolumn{2}{c}{4}& \multicolumn{2}{c}{44}& \multicolumn{2}{c}{28}& \multicolumn{2}{c}{25}& \multicolumn{2}{c}{11}& \multicolumn{2}{c}{122}\\
\bottomrule
\end{tabular}
}
}
\label{tab:yields_obs}
\end{table}
 
A signal strength of \obsmu\ is measured with respect to the {\sherpa} fiducial cross section prediction for \sswwjj\ electroweak production, where the systematic uncertainty also includes the absolute normalization uncertainty of this prediction. This corresponds to a fiducial signal cross section of:
\begin{equation*}
\obsxsec,
\end{equation*}
where the uncertainties correspond to the statistical, experimental systematic, theory modeling systematic and luminosity uncertainties, respectively. The experimental systematic uncertainty includes the detector systematic uncertainties and the uncertainties in estimating all background processes except for the $\wzjj$ and $\sswwjj$ strong production processes that are accounted for in the modeling systematic uncertainty. Table~\ref{tabaux:systematics} summarizes the impacts of different components of systematic uncertainty.
 
The measured fiducial cross section includes contributions from both the $\sswwjj$ electroweak production and its interference with the $\sswwjj$ strong production, estimated to be approximately 6\% of the predicted fiducial cross section for $\sswwjj$ electroweak production. The fiducial cross section for the $\sswwjj$ electroweak production, without the interference effect, is predicted by {\sherpa} and {\powheg}+{\pythia} to be $\shpxsec$ and $\pbpxsec$, respectively. The impact on the measured fiducial cross section of using {\powheg}+{\pythia} instead of {\sherpa} to generate the {\mjj} signal template was tested and found to be smaller than the $3.6\%$ signal modeling uncertainty.
 
\begin{table}[htb]
\caption{Impact of different components of systematic uncertainty on the measured fiducial cross section, without taking into account correlations. The impact of one source of systematic uncertainty is computed by first performing the fit with the corresponding nuisance parameter fixed to one standard deviation up or down from the value obtained in the nominal fit, then these up and down variations are symmetrized. The impacts of several sources of systematic uncertainty are added in quadrature for each component. The categorization of sources of systematic uncertainties into experimental and theory modeling correspond to those used for the measured fiducial cross section.
}
\begin{center}
\begin{tabular}{p{2ex}lc} 
\toprule 
\multicolumn{2}{l}{Source} & Impact [\%] \\ 
\midrule 
\multicolumn{2}{l}{Experimental}           & \\ 
\midrule 
& Electron energy scale and resolution, and efficiency          & ${\hphantom{0}0.6}$ \\ 
& Muon momentum scale and resolution, and efficiency            & ${\hphantom{0}1.3}$ \\ 
& Jet energy and $E_{\mathrm{T}}^{\mathrm{miss}}$ scale and resolution & ${\hphantom{0}3.2}$ \\ 
& $b$-tagging inefficiency                & ${\hphantom{0}2.1}$ \\ 
& Pileup modeling                         & ${\hphantom{0}1.6}$ \\ 
& Background, statistical & ${\hphantom{0}3.2}$ \\ 
& Background, misid.\ leptons  & ${\hphantom{0}3.3}$ \\ 
& Background, charge misrec.   & ${\hphantom{0}0.3}$ \\ 
& Background, other & ${\hphantom{0}1.8}$ \\ 
\midrule 
\multicolumn{2}{l}{Theory modeling}           & \\ 
\midrule 
& \sswwjj\ electroweak-strong interference & ${\hphantom{0}1.0}$ \\ 
& \sswwew, EW corrections & ${\hphantom{0}1.4}$ \\ 
& \sswwew, shower, scale, PDF \& $\alpha_s$ & ${\hphantom{0}2.8}$ \\ 
& \sswwqcd         & ${\hphantom{0}2.9}$ \\ 
& $WZ$             & ${\hphantom{0}3.3}$ \\ 
\midrule 
\multicolumn{2}{l}{Luminosity}           & ${\hphantom{0}2.4}$ \\ 
\bottomrule 
\end{tabular} 
\label{tabaux:systematics}
\end{center}
\end{table}
 
\FloatBarrier
In summary, the electroweak $VVjj$ production process was studied in the \sswwjj\ final state using {\lumi} of $pp$ collision data recorded at $\sqrt{s}=13$~\TeV\ by the ATLAS detector at the LHC. This process includes VBS diagrams which probe the mechanism of electroweak symmetry breaking. An excess of events is observed and the background-only hypothesis is rejected with a significance of 6.5$\sigma$. The fiducial cross section for \sswwjj\ electroweak production is measured to be $\obsxsec$.
 
\section*{Acknowledgments}
 
We thank CERN for the very successful operation of the LHC, as well as the
support staff from our institutions without whom ATLAS could not be
operated efficiently.
 
We acknowledge the support of ANPCyT, Argentina; YerPhI, Armenia; ARC, Australia; BMWFW and FWF, Austria; ANAS, Azerbaijan; SSTC, Belarus; CNPq and FAPESP, Brazil; NSERC, NRC and CFI, Canada; CERN; CONICYT, Chile; CAS, MOST and NSFC, China; COLCIENCIAS, Colombia; MSMT CR, MPO CR and VSC CR, Czech Republic; DNRF and DNSRC, Denmark; IN2P3-CNRS, CEA-DRF/IRFU, France; SRNSFG, Georgia; BMBF, HGF, and MPG, Germany; GSRT, Greece; RGC, Hong Kong SAR, China; ISF, I-CORE and Benoziyo Center, Israel; INFN, Italy; MEXT and JSPS, Japan; CNRST, Morocco; NWO, Netherlands; RCN, Norway; MNiSW and NCN, Poland; FCT, Portugal; MNE/IFA, Romania; MES of Russia and NRC KI, Russian Federation; JINR; MESTD, Serbia; MSSR, Slovakia; ARRS and MIZ\v{S}, Slovenia; DST/NRF, South Africa; MINECO, Spain; SRC and Wallenberg Foundation, Sweden; SERI, SNSF and Cantons of Bern and Geneva, Switzerland; MOST, Taiwan; TAEK, Turkey; STFC, United Kingdom; DOE and NSF, United States of America. In addition, individual groups and members have received support from BCKDF, the Canada Council, CANARIE, CRC, Compute Canada, FQRNT, and the Ontario Innovation Trust, Canada; EPLANET, ERC, ERDF, FP7, Horizon 2020 and Marie Sk{\l}odowska-Curie Actions, European Union; Investissements d'Avenir Labex and Idex, ANR, R{\'e}gion Auvergne and Fondation Partager le Savoir, France; DFG and AvH Foundation, Germany; Herakleitos, Thales and Aristeia programmes co-financed by EU-ESF and the Greek NSRF; BSF, GIF and Minerva, Israel; BRF, Norway; CERCA Programme Generalitat de Catalunya, Generalitat Valenciana, Spain; the Royal Society and Leverhulme Trust, United Kingdom.
 
The crucial computing support from all WLCG partners is acknowledged gratefully, in particular from CERN, the ATLAS Tier-1 facilities at TRIUMF (Canada), NDGF (Denmark, Norway, Sweden), CC-IN2P3 (France), KIT/GridKA (Germany), INFN-CNAF (Italy), NL-T1 (Netherlands), PIC (Spain), ASGC (Taiwan), RAL (UK) and BNL (USA), the Tier-2 facilities worldwide and large non-WLCG resource providers. Major contributors of computing resources are listed in Ref.~\cite{ATL-GEN-PUB-2016-002}.

\printbibliography
 
\clearpage
 
\begin{flushleft}
{\Large The ATLAS Collaboration}

\bigskip

M.~Aaboud$^\textrm{\scriptsize 35d}$,    
G.~Aad$^\textrm{\scriptsize 100}$,    
B.~Abbott$^\textrm{\scriptsize 127}$,    
O.~Abdinov$^\textrm{\scriptsize 13,*}$,    
B.~Abeloos$^\textrm{\scriptsize 131}$,    
D.K.~Abhayasinghe$^\textrm{\scriptsize 92}$,    
S.H.~Abidi$^\textrm{\scriptsize 166}$,    
O.S.~AbouZeid$^\textrm{\scriptsize 40}$,    
N.L.~Abraham$^\textrm{\scriptsize 155}$,    
H.~Abramowicz$^\textrm{\scriptsize 160}$
% ...
\bigskip
\\

% ...
$^{13}$Institute of Physics, Azerbaijan Academy of Sciences, Baku; Azerbaijan.\\
% ...
$^{35}$$^{(a)}$Facult\'e des Sciences Ain Chock, R\'eseau Universitaire de Physique des Hautes Energies - Universit\'e Hassan II, Casablanca;$^{(b)}$Facult\'{e} des Sciences, Universit\'{e} Ibn-Tofail, K\'{e}nitra;$^{(c)}$Facult\'e des Sciences Semlalia, Universit\'e Cadi Ayyad, LPHEA-Marrakech;$^{(d)}$Facult\'e des Sciences, Universit\'e Mohamed Premier and LPTPM, Oujda;$^{(e)}$Facult\'e des sciences, Universit\'e Mohammed V, Rabat; Morocco.\\
% ...
$^{40}$Niels Bohr Institute, University of Copenhagen, Copenhagen; Denmark.\\
% ...
$^{92}$Department of Physics, Royal Holloway University of London, Egham; United Kingdom.\\
% ...
$^{100}$CPPM, Aix-Marseille Universit\'e, CNRS/IN2P3, Marseille; France.\\
% ...
$^{127}$Homer L. Dodge Department of Physics and Astronomy, University of Oklahoma, Norman OK; United States of America.\\
% ...
$^{131}$LAL, Universit\'e Paris-Sud, CNRS/IN2P3, Universit\'e Paris-Saclay, Orsay; France.\\
% ...
$^{155}$Department of Physics and Astronomy, University of Sussex, Brighton; United Kingdom.\\
% ...
$^{160}$Raymond and Beverly Sackler School of Physics and Astronomy, Tel Aviv University, Tel Aviv; Israel.\\
% ...
$^{166}$Department of Physics, University of Toronto, Toronto ON; Canada.\\
% ...
$^{a}$ Also at Borough of Manhattan Community College, City University of New York, New York NY; United States of America.\\
$^{b}$ Also at Centre for High Performance Computing, CSIR Campus, Rosebank, Cape Town; South Africa.\\
$^{c}$ Also at CERN, Geneva; Switzerland.\\
$^{d}$ Also at CPPM, Aix-Marseille Universit\'e, CNRS/IN2P3, Marseille; France.\\
$^{e}$ Also at D\'epartement de Physique Nucl\'eaire et Corpusculaire, Universit\'e de Gen\`eve, Gen\`eve; Switzerland.\\
% ...
$^{*}$ Deceased

\end{flushleft}

 
\end{document}